\documentclass[sigconf]{acmart}
\usepackage[binary-units,per-mode=symbol,per-symbol=p]{siunitx}
\usepackage[utf8]{inputenc}
\usepackage{adjustbox}
\usepackage[ruled,vlined]{algorithm2e}
\usepackage{algorithmic}
\usepackage{graphicx}
\usepackage{textcomp}
\usepackage{xcolor}
\usepackage{balance}
\usepackage{enumitem}
\usepackage{setspace}
\usepackage{wrapfig}
\usepackage{listings}
\usepackage{color}
\usepackage{caption}
\usepackage{subcaption}
\usepackage{tikz}
\usetikzlibrary{shapes.geometric, arrows}
\usetikzlibrary{fit}
\usetikzlibrary{calc}
\usepackage{siunitx}

\tikzstyle{process_g} = [rectangle, minimum width=2.3cm, minimum height=0.75cm, text centered, text width=2.3cm, draw=black, fill=gray!40]

\tikzstyle{process} = [rectangle, minimum width=2.3cm, minimum height=0.75cm, text centered, text width=2.3cm, draw=black, fill=gray!10]
\tikzstyle{arrow} = [thick,->,>=stealth]

\usepackage{lipsum} 
\usepackage{hyperref}
\definecolor{lime}{HTML}{A6CE39}
\DeclareRobustCommand{\orcidicon}{%
	\begin{tikzpicture}
	\draw[lime, fill=lime] (0,0) 
	circle [radius=0.16] 
	node[white] {{\fontfamily{qag}\selectfont \tiny ID}};
	\draw[white, fill=white] (-0.0625,0.095) 
	circle [radius=0.007];
	\end{tikzpicture}
	\hspace{-2mm}
}

\foreach \x in {A, ..., Z}{%
	\expandafter\xdef\csname orcid\x\endcsname{\noexpand\href{https://orcid.org/\csname orcidauthor\x\endcsname}{\noexpand\orcidicon}}
}

\newcommand{\cf}{\emph{cf.}\xspace}
\newcommand{\etal}{\emph{et al.}\xspace}
\newcommand{\ie}{\emph{i.e.}, }

\newcommand{\etc}{\emph{etc. }}
\newcommand{\HAS}{\emph{HTTP Adaptive Streaming }}

\newcommand{\DCT}{\emph{Discrete Cosine Transform }}

\newcommand{\VQA}{{video quality assessment}}

\newcommand{\FRVQA}{\texttt{FR-VQA}\xspace}
\newcommand{\RRVQA}{\texttt{RR-VQA}\xspace}

\newcommand{\rrtif}{\texttt{VQ-TIF}\xspace}

\newcommand{\pcc}{\emph{Pearson Correlation Coefficient}\xspace}
\newcommand{\lstm}{\emph{long short-term memory}\xspace}

\newcommand{\EY}{$E_{\text{Y}}$\xspace}
\newcommand{\LY}{$L_{\text{Y}}$\xspace}
\newcommand{\h}{$h$\xspace}

\copyrightyear{2024}
\acmYear{2024}
\setcopyright{rightsretained}
\acmConference[MHV '24]{Mile-High Video Conference}{February 11--14, 2024}{Denver, CO, USA}
\acmBooktitle{Mile-High Video Conference (MHV '24), February 11--14, 2024, Denver, CO, USA}
\acmDOI{10.1145/3638036.3640798}
\acmISBN{979-8-4007-0493-2/24/02}

\begin{document}

\title{Video Quality Assessment with Texture Information Fusion for Streaming Applications}

\author{Vignesh V Menon}
\email{vignesh.menon@hhi.fraunhofer.de}
\orcid{0000-0003-1454-6146}
\affiliation{
  \institution{\small{Video Communication and Applications Dept}}
  \institution{Fraunhofer HHI}
  \city{Berlin}
  \country{Germany}
}

\author{Prajit T Rajendran}
\email{prajit.thazhurazhikath@cea.fr}
\orcid{0000-0002-8283-9891}
\affiliation{
  \institution{\small{CEA, List, F-91120 Palaiseau}}
  \institution{Université Paris-Saclay}
  \city{Paris}
  \country{France}
}

\author{Reza Farahani}
\email{reza.farahani@aau.at}
\orcid{0000-0002-2376-5802}
\affiliation{
  \institution{\small{Institute of Information Technology}}
  \institution{Alpen-Adria-Universität Klagenfurt}
  \city{Klagenfurt}
  \country{Austria}
}

\author{Klaus Schoeffmann}
\email{klaus.schoeffmann@aau.at}
\affiliation{
  \institution{\small{Institute of Information Technology}}
  \institution{Alpen-Adria-Universität}
  \city{Klagenfurt}
  \country{Austria}
}

\author{Christian Timmerer}
\email{christian.timmerer@aau.at}
\orcid{0000-0002-0031-5243}
\affiliation{
  \institution{\small{Institute of Information Technology}}
  \institution{Alpen-Adria-Universität}
  \city{Klagenfurt}
  \country{Austria}
}

\renewcommand{\shortauthors}{Vignesh V Menon~\etal}

\begin{abstract}
The rise in video streaming applications has increased the demand for \VQA~(VQA). In 2016, Netflix introduced \textit{Video Multi-Method Assessment Fusion} (VMAF), a full reference VQA metric that strongly correlates with perceptual quality, but its computation is time-intensive. We propose a \DCT~(DCT)-energy-based VQA with texture information fusion (\rrtif) model for video streaming applications that determines the visual quality of the reconstructed video compared to the original video. \rrtif extracts Structural Similarity (SSIM) and spatiotemporal features of the frames from the original and reconstructed videos and fuses them using a \lstm~(LSTM)-based model to estimate the visual quality. Experimental results show that \rrtif estimates the visual quality with a \textit{Pearson Correlation Coefficient} (PCC) of \SI{0.96}{} and a \textit{Mean Absolute Error} (MAE) of \SI{2.71}{}, on average, compared to the ground truth VMAF scores. Additionally, \rrtif estimates the visual quality at a rate of \SI{9.14}{} times faster than the state-of-the-art VMAF implementation, along with an \SI{89.44}{\percent} reduction in energy consumption, assuming an Ultra HD (2160p) display resolution.
\end{abstract}




\maketitle

\section{Introduction}
\begin{figure}[t]
\centering
\includegraphics[width=0.355\textwidth]{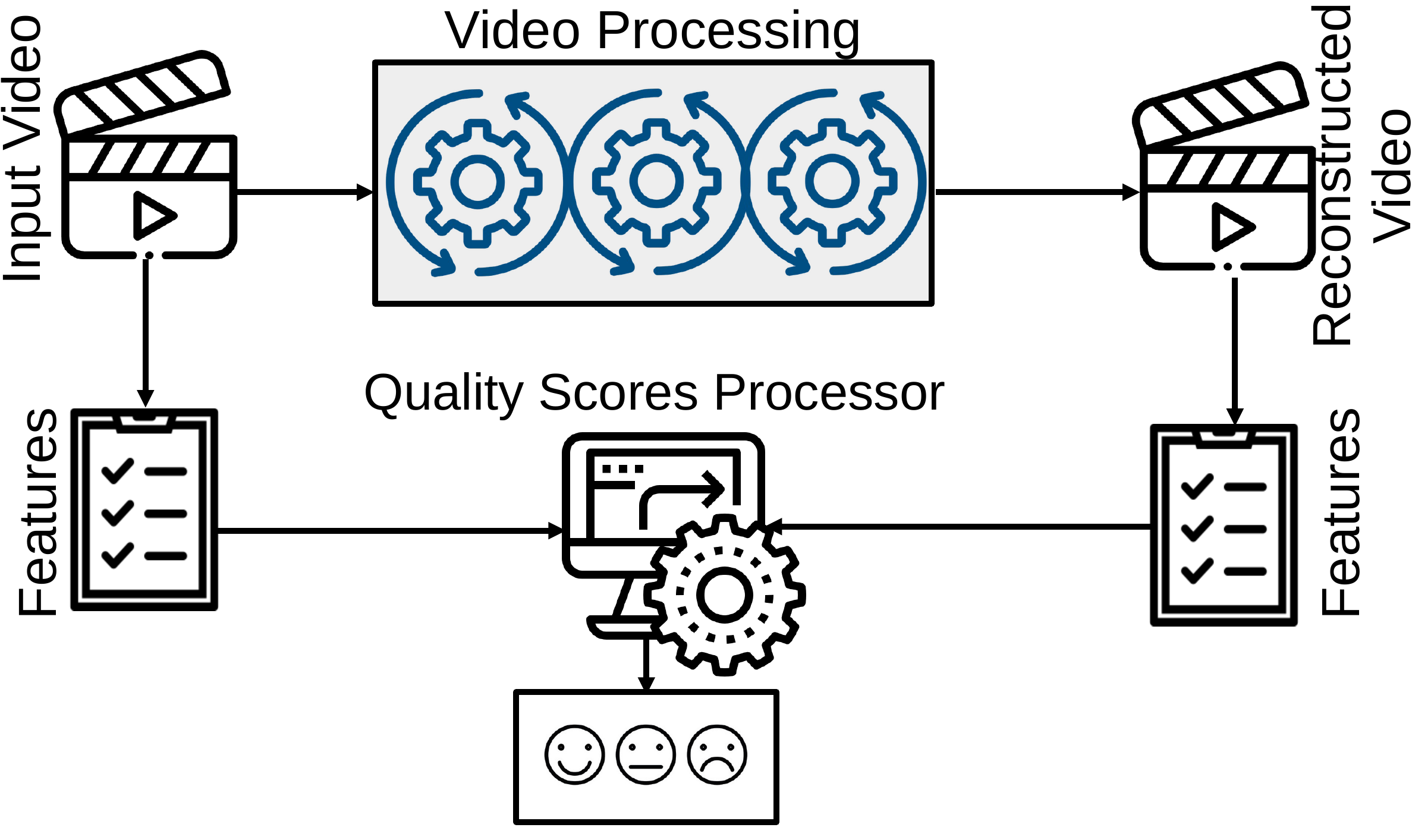}
\caption{The structure of state-of-the-art \RRVQA methods utilized, especially within streaming video coding systems.}
\label{fig:vqa_sota}
\end{figure}

With the ever-increasing demands for high-definition video streaming services, the need for \VQA (VQA) is growing potentially. VQA plays an essential role in video processing from capturing to rendering, including compression, transmission, restoration, and display~\cite{VQ_survey}. With all the available encoding options and trade-offs to consider in \HAS (HAS)~\cite{DASH_Survey}, having a lightweight and reliable VQA method is crucial. According to the degree of information available for the reference video signals, VQA is classified into \textit{full reference} (FR), \textit{reduced reference} (RR), and \textit{no reference} (NR) methods. NR-VQA methods are ``blind'' where the original video content is not used for the quality assessment, leading to an unreliable VQA~\cite{VQ_survey}. The advantage of \RRVQA lies in its ability to evaluate video quality using limited information, making it more suitable for real-time VQA, especially in adaptive streaming or live broadcast scenarios~\cite{mcbe_ref, cvfr_ref}. As shown in Figure~\ref{fig:vqa_sota}, \RRVQA models often focus on extracting specific features or subsets of data, such as selected frames, segments, or critical information related to video content. This approach allows \RRVQA to provide quality evaluations even when access to the entire reference is restricted or when performing evaluations in bandwidth-constrained environments~\cite{rr_vqa_survey}. However, the downside of \RRVQA is that it might not provide the same accuracy or granularity in the quality evaluation as \FRVQA due to the limited information it uses. On the other hand, \FRVQA can offer a more detailed comparison between the distorted video and the entire reference, leading to a more precise VQA.

\textit{Peak Signal to Noise Ratio} (PSNR) continues to be the predominant industry benchmark for standardizing video codecs. PSNR is an effective method to generate a numeric value that compares an original input file and a coded output file. The limitations of PSNR are \textit{(i)} its failure to account for the temporal nature of compression artifacts and \textit{(ii)} lack of correlation between PSNR improvements and subjective quality, particularly in the presence of camera noise~\cite{psnr_ref1,psnr_ref2}. \textit{Structural Similarity} (SSIM) is another image quality metric introduced in 2004~\cite{ref1} that considers image degradation as a perceived change in structural information. It also incorporates critical perceptual phenomena, including both luminance masking and contrast masking terms. \textit{Video Multi-Method Assessment Fusion} (VMAF) was explicitly formulated by Netflix to correlate strongly with subjective Mean Opinion Scores (MOS). Using machine learning techniques, a large sample of MOS was used as ground truth to train a quality estimation model. Among the state-of-the-art VQA metrics, VMAF achieves the highest correlation with the \textit{Difference Mean Opinion Score} (DMOS). However, VMAF is computationally intensive, and its time complexity is very high compared to the PSNR and SSIM metrics~\cite{vmaf_ref_madhu}. 

We estimate visual quality to support low-latency VQA in video streaming applications. The expected computation time should be comparable to PSNR and SSIM, with the highest possible accuracy compared to the VMAF score. The proposed VQA method is expected to replace the state-of-the-art VMAF computation in streaming applications.  

\textit{\textbf{Contributions:}} The key contributions of this paper are as follows:
\begin{enumerate}
\item We propose a fast machine learning-based \RRVQA model using texture information fusion (\rrtif), which is implemented in real-time to determine visual quality. To our knowledge, this is among the first efforts on \RRVQA implementation of VMAF estimation.
\item \rrtif encompasses \DCT~(DCT)-energy-based texture information and SSIM fusion between the original and the reconstructed video. It extracts video complexity features like brightness and spatiotemporal texture information from the luma channels of the videos. The extracted features are fused using an \lstm (LSTM)-based model to determine the \rrtif score.
\item The \rrtif score is estimated at a rate of \SI{9.14}{} times faster than the state-of-the-art VMAF evaluation, with a \pcc (PCC) of \SI{0.96}{} with the ground truth VMAF values and \SI{89.44}{\percent} reduction in energy consumption, assuming a UHD (2160p) display.
\end{enumerate}

The paper has six sections. Section~\ref{sota} analyzes VQA state-of-the-art works. Section~\ref{RL} reviews the related work. We elaborate on the details of the \rrtif model in Section~\ref{sec:rrtif_framework}. Section~\ref{sec:evaluation} explains the evaluation setup and experimental results before concluding the paper in Section~\ref{sec:conclusion_future_dir}.
\section{State-of-the-art VQA Analysis}\label{sota}
\paragraph{Peak Signal-to-Noise Ratio (PSNR)} is a conventional quality metric used in video quality assessment due to its simplicity and ease of computation~\cite{psnr_ref1,psnr_ref2}. Its primary advantage lies in its straightforwardness and popularity as a benchmark for measuring compression performance, which helps to compare different encoding methods or qualities. However, in the context of adaptive streaming, PSNR has several drawbacks. It often does not correlate well with perceived visual quality, especially at lower bitrates or scenarios with complex video content~\cite{vmaf_comp}. Moreover, PSNR fails to capture perceptual differences and remains insensitive to human visual perception. This renders it unsuitable for evaluating artifacts or distortions that may be visually noticeable but are not adequately represented in PSNR values. In addition, its performance diminishes when evaluating video quality affected by compression artifacts or when the video undergoes format or resolution changes. Consequently, in adaptive streaming scenarios where subjective user experience is paramount, relying solely on PSNR can lead to suboptimal quality decisions and hinder delivering an optimal quality of experience (QoE) to viewers~\cite{jtps_ref}.

\paragraph{Structural similarity (SSIM)} considers structural information, mimicking human visual perception more closely than PSNR~\cite{ref1}. It correlates better with perceived quality changes, particularly in compression, noise, or distortion scenarios~\cite{ssim_ref2,ssim_ref3}. This makes SSIM more effective in capturing subtle variations in video quality that can impact viewer experience, particularly at lower bitrates. SSIM also has limitations in adaptive streaming contexts. It can be sensitive to specific distortions and may not consistently reflect human perception across all video content~\cite{vmaf_comp}. Moreover, its performance can vary depending on the complexity of the content and the types of distortions present in the video. It also does not always align with subjective assessments and may not accurately represent viewers' visual quality. As a result, while SSIM provides more nuanced insights than PSNR, it is not the only comprehensive metric for assessing video quality in adaptive streaming settings, where a holistic evaluation considering multiple metrics and subjective perception is crucial.

\paragraph{Video Multi-Method Assessment Fusion (VMAF)} is a full-reference, perceptual video quality metric that aims to approximate human perception of video quality. This metric is focused on quality degradation due to compression and rescaling. VMAF estimates the perceived quality score by computing scores from multiple quality assessment algorithms and fusing them with a support vector machine. In contrast to PSNR and SSIM metrics, which do not take temporal information into account~\cite{ssim_ref3}, three image fidelity metrics and one temporal signal have been chosen as features of the SVM: \textit{(i)} Antinoise SNR, \textit{(ii)} Detail Loss Measure, \textit{(iii)} Visual Information Fidelity, and \textit{(iv)} Mean Co-Located Pixel Difference (MCPD). An essential feature is the MCPD of a frame to the previous frame (\ie the temporal component). A VMAF score is more straightforward to understand because it operates in a linear range of \SI{0}to \SI{100}, whereas PSNR and SSIM are logarithmic. It considers scaling and compression artifacts and has a model trained for mobile video consumption~\cite{vmaf_comp}.

\begin{table}[t]
\caption{Pearson Correlation of VQA metrics.}
\centering
\resizebox{0.505\linewidth}{!}{
\begin{tabular}{l|c|c|c}
\specialrule{.12em}{.05em}{.05em}
\specialrule{.12em}{.05em}{.05em}
Metric & PSNR & SSIM & VMAF \\
\specialrule{.12em}{.05em}{.05em}
\specialrule{.12em}{.05em}{.05em}
PSNR & 1.00 & 0.70 & 0.83 \\
SSIM & 0.70 & 1.00 & 0.88 \\
VMAF & 0.83 & 0.88 & 1.00 \\
\specialrule{.12em}{.05em}{.05em}
\specialrule{.12em}{.05em}{.05em}
\end{tabular}}
\label{tab:pcc_vqa}
\end{table}

\begin{figure*}[t]
\centering
\begin{subfigure}{0.323\textwidth}
    \centering
    \includegraphics[width=\textwidth]{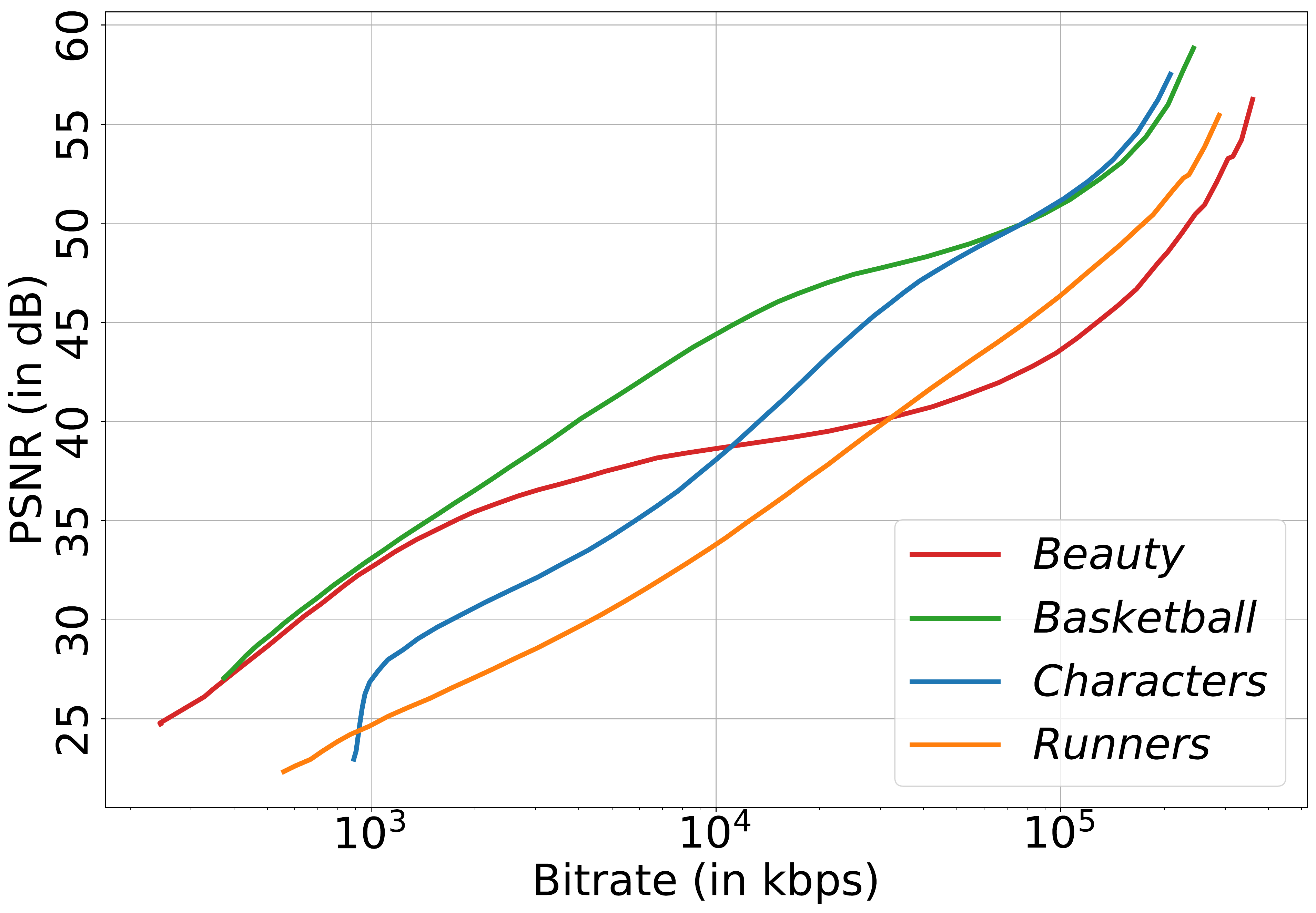}
    \caption{PSNR}
        \label{fig:rd_psnr}
\end{subfigure}
\hfill
\begin{subfigure}{0.323\textwidth}
    \centering
    \includegraphics[width=\textwidth]{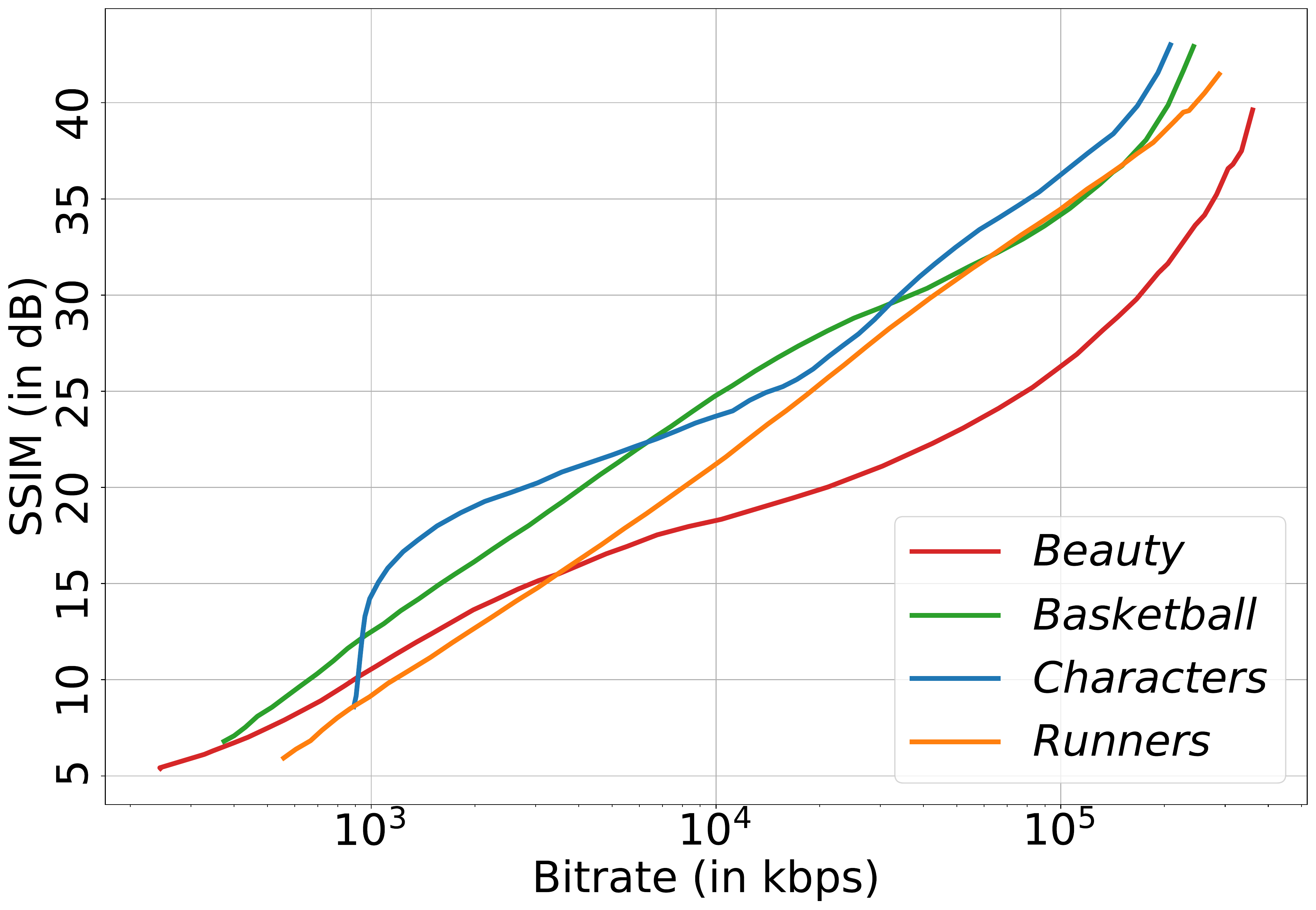}
    \caption{SSIM}
        \label{fig:rd_ssim}    
\end{subfigure}
\hfill
\begin{subfigure}{0.323\textwidth}
    \centering
    \includegraphics[width=\textwidth]{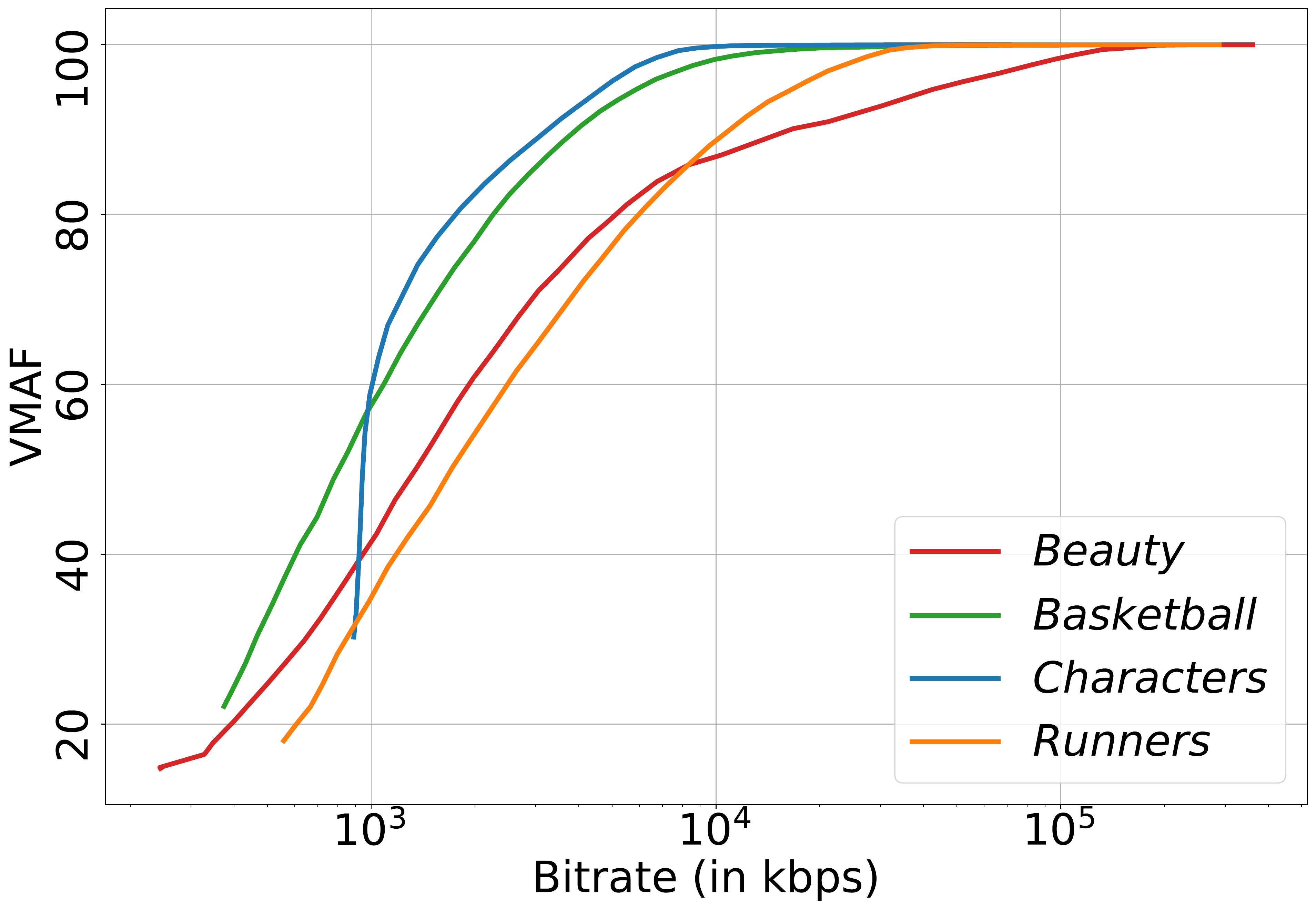}
    \caption{VMAF}
    \label{fig:rd_vmaf}    
\end{subfigure}
\caption{Rate-distortion (RD) curves of selected segments of different spatiotemporal complexities -- \textit{Beauty} (\EY=59.90, \h=17.49, \LY=89.25), \textit{Basketball} (\EY=15.30, \h=12.59, \LY=108.18), \textit{Characters} (\EY=45.42, \h=36.88, \LY=134.56), and \textit{Runners} (\EY=105.85, \h=22.48, \LY=126.60). The segments are downsampled to 30\,fps and encoded with the x264 AVC encoder using \textit{ultrafast} preset and CRF rate control.}
\label{fig:rd_plots}
\end{figure*}

Table~\ref{tab:pcc_vqa} shows the Pearson Correlation of PSNR, SSIM, and VMAF quality metrics analyzed for a thousand video sequences from the Inter4k Dataset~\cite{inter4k_ref}. These sequences were encoded at Ultra High Definition (2160p) resolution using the x264 AVC encoder, employing the \textit{ultrafast} preset and constant rate factor (CRF). CRF values ranging between \SI{1} and \SI{51}{} are used in the analysis. The correlation of the VMAF score with the PSNR and SSIM scores is \SI{0.83}{} and \SI{0.88}{}, respectively. The correlation can also be observed graphically in Figure~\ref{fig:rd_plots}, which demonstrates the rate-distortion (RD) curves of selected video sequences from UVG~\cite{uvg_ref}, MCML~\cite{mcml_video_ref}, and SJTU~\cite{sjtu_video_ref} datasets, based on their spatio-temporal complexity, where distortion is measured using PSNR, SSIM, and VMAF.

\section{Related work}\label{RL}
\RRVQA approaches are categorized into three primary types based on the nature of the features they utilize: \textit{(i)} pixel-based, \textit{(ii)} frequency-based, and \textit{(iii)} bitstream-based methods.

\paragraph{Pixel-based \RRVQA:} This approach involves extracting and analyzing spatial or temporal pixel-level information from distorted and reference video frames. It typically involves comparing pixel-wise differences or using metrics derived from pixel values, such as MSE or SSIM. Pixel-based methods directly assess visual discrepancies at the pixel level, often considering factors such as luminance, color, and spatial arrangement~\cite{rr_vqa1,rr_vqa2}.

\paragraph{Frequency-based \RRVQA:} Thie method is based on the analysis of video content in the frequency domain, mainly using information derived from transformations such as the discrete cosine transform (DCT) \cite{rr_dct1} or the discrete wavelet transform (DWT)~\cite{dwt_rrvqa}. By examining the frequency components and their differences between the reference and distorted video, this scheme captures variations in specific frequency bands or coefficients, providing insights into how the signal's frequency distribution influences perceptual quality.

\paragraph{Bitstream-based \RRVQA:} In this approach, the video bitstream data is analyzed to extract features without decoding the video content. It involves inspecting parameters or metadata within the compressed bitstream, including coding information, motion vectors, quantization parameters, or syntax elements~\cite{bitstream_vqa_ref1}. By evaluating these aspects, bitstream-based methods aim to infer quality differences without requiring full access to the original video content.

Other than the aforementioned types of \RRVQA, our previous work~\cite{tqpm_ref} introduced a reduced reference transcoding quality prediction model (TQPM) to determine the VMAF of the video possibly transcoded in multiple stages. The quality is predicted using video complexity features (\ie the video’s brightness, spatial texture information, and temporal activity) and the target bitrate representation of each transcoding stage.

Since PSNR remains the de-facto industry standard for video quality evaluation, many \RRVQA methods are developed to evaluate it~\cite{rr_psnr1,rr_psnr2}. Furthermore, there are methods that predict SSIM~\cite{rr_ssim1, rr_ssim2}, Spatiotemporal RR Entropic Differences (STRRED)~\cite{rr_strred1}, and Spatial RR Entropic Differences (SRRED)~\cite{rr_srred1} metrics. However, the described metrics have limitations, such as neglecting the temporal nature of compression artifacts~\cite{vmaf_ref1}. Each \RRVQA category has advantages and limitations based on the type of features extracted and the complexities involved in assessing perceptual video quality. The choice of method often depends on the specific application, available resources, and the extent to which the method aligns with the perceptual attributes most relevant to the evaluation task. Although \RRVQA methods are implemented to estimate PSNR and SSIM, no work in the literature determines VMAF, although VMAF yields VQA closer to the visual perception of HVS than its counterparts. However, it is significantly time-consuming compared to other metrics.
\begin{figure*}[t]
\centering
\includegraphics[width=0.9\linewidth]{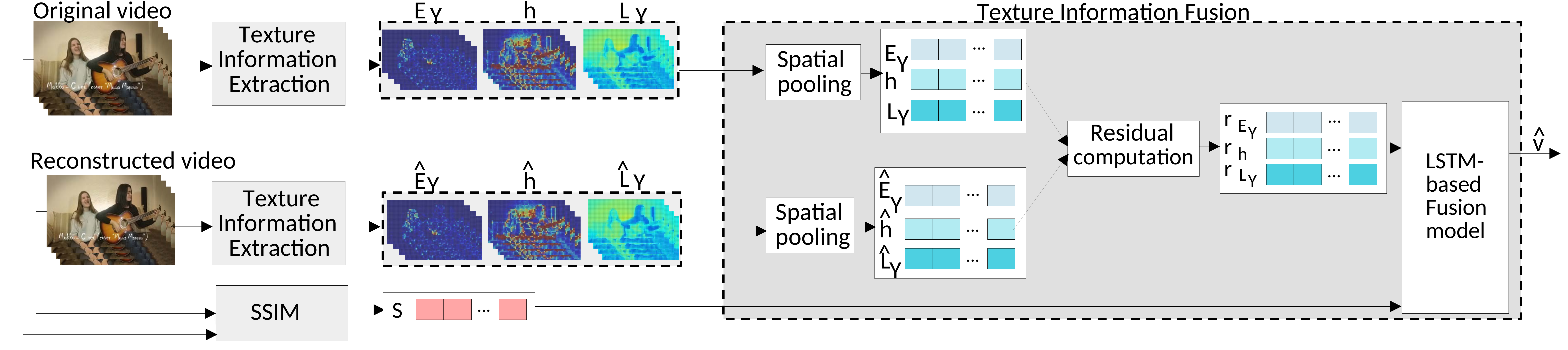}
\caption{VQA for a video segment using \rrtif model envisioned in this paper.}
\label{fig:rrtif_arch}
\end{figure*}
\section{\rrtif model}
\label{sec:rrtif_framework}
The architecture of the proposed \rrtif-based VMAF estimation is illustrated in Figure~\ref{fig:rrtif_arch}. Since the correlation between SSIM and VMAF is very high (as observed in Table~\ref{tab:pcc_vqa}), and the computation time of SSIM is significantly lower than VMAF, we select SSIM as a main feature to compute VMAF. In this architecture, the input video segment is divided into multiple chunks. The main characteristics of \rrtif include:
\begin{enumerate}
\item frame-wise \textit{texture information extraction} for each chunk (Section~\ref{sec:feature_extraction})
\item SSIM calculation
\item \textit{texture information fusion}, where the features and the computed SSIM are fused using an LSTM-based model to determine the \rrtif score for each chunk (Section~\ref{sec:tif})
\end{enumerate}
The \rrtif scores obtained for each chunk are averaged as the VMAF for the reconstructed video segment.
\subsection{Texture Information Extraction}
\label{sec:feature_extraction}
An intuitive feature extraction method would be utilizing Convolutional Neural Networks (CNNs)~\cite{3d_cnn_vqa_ref}. However, such models have several inherent disadvantages, such as a longer training time, more inference time, and storage requirements, making them impractical for streaming scenarios. Although CNN-based approaches could produce rich features, simpler models that yield significant prediction performance are more suitable for video streaming applications. The popular state-of-the-art video complexity features are Spatial Information (SI) and Temporal Information (TI)~\cite{siti_itu_ref}. However, the correlation of SI and TI features with the encoding output features such as bitrate, encoding time, \etc are very low, which is insufficient for encoding parameter prediction in streaming applications~\cite{vca_ref}.

In this paper, three DCT-energy-based features~\cite{dct_ref}, the average luma texture energy \EY, the average gradient of the luma texture energy \h, and the average luminescence \LY are used as the texture information measures~\cite{vca_ref, jtps_ref}.
\subsection{Texture Information Fusion}
\label{sec:tif}
The texture information fusion step of \rrtif is accomplished using the following steps:

\subsubsection{Spatial pooling} 
The video segments are divided into $T$ chunks with a fixed number of frames (\ie $f_c$). The averages of the \EY, \h, and \LY features of each frame in the chunk are calculated to obtain the spatially pooled representation of the chunk, expressed as:
$X = \{x_{1}, x_{2},.., x_{f_c}\}$, and  $\hat{X} = \{\hat{x}_{1}, \hat{x}_{2},.., \hat{x}_{f_c}\}$, where, $x_{i}$ and $\hat{x}_{i}$ are the $i^{th}$ frame feature set associated to the original and reconstructed video chunks, respectively.
\begin{align}
x_{i} &= [E_{i}, h_{i}, L_{i}],\\
\hat{x}_{i} &= [\hat{E}_{i}, \hat{h}_{i}, \hat{L}_{i}] \hspace{0.4cm} \forall i \in [1,{f_c}]
\end{align}
\subsubsection{Residual computation}
Residual features are formed by subtracting the original video texture information features from the reconstructed video features. This difference is known as the error or residual feature, expressed as:
\begin{align}
    r_{E_{i}} = E_{i} - \hat{E}_{i}\\
    r_{h_{i}} = h_{i} - \hat{h}_{i}\\
    r_{L_{i}} = L_{i} - \hat{L}_{i}
\end{align}
where $i \in [1,{f_c}]$. The residual features usually have low information entropy, as the original and reconstructed video frames are similar. The entropy increases with increased distortion introduced in the reconstructed video.


\subsubsection{Fusion}
The fusion of the texture information features is established using a \lstm (LSTM). LSTM is selected as a model for processing sequential data, making it suitable for combining information for temporally adjacent frames in a video. An advantage of LSTM models is better handling of long-term dependencies in long sequences. Each chunk's feature averages are considered separate data points in the model training process.
Therefore, the input data consists of the residuals of the spatially pooled luma texture information features extracted per frame of the video chunk. Moreover, frame-wise SSIM values denoted by $S = \{s_{1}, s_{2},.., s_{f_c}\}$ are appended to the residual features. The prediction model is a function of the residual features of the frames and the SSIM values in a chunk, as shown in Eq.~\ref{eq:lstm_in}. This approach can fuse feature information from temporally adjacent frames to estimate visual quality. 
\begin{equation}
\label{eq:lstm_in}
\Tilde{x_{i}} = [r_{i} | s_{i}]^{T} \hspace{0.4cm} i \in [1, {f_c}]
\end{equation}
where $r_{i} = [r_{E_{i}}, r_{h_{i}}, r_{L_{i}}]$. The estimated \rrtif score per chunk $\hat{v}$ can be presented as: $\hat{v} = f(\Tilde{x})$. The \rrtif score of the reconstructed video segment is the average of the $\hat{v}$ values estimated for every chunk.
\subsection{\rrtif Implementation}
The \EY, \h, and \LY features of the original and reconstructed video segments are extracted using VCA v2.0 open-source video complexity analyzer~\cite{vca_ref}. LSTM model is implemented using the Keras~\cite{keras_ref} machine learning framework. LSTMs can address the vanishing gradient problem, allowing them to retain information over extended intervals. The input shape of the network is $[f_{c} \times 4]$. In the network architecture implemented in this paper, there are two layers; the first layer consists of \SI{200}{} LSTM cells, which serve as the memory hub of the network, empowering the model to discern and remember long-range dependencies in input sequences. A dense layer follows this, as the target variable consists of a single value. The dual-layer composition enhances the model's capacity to navigate and comprehend intricate sequential data while ensuring robust performance in regression-oriented tasks. The loss function and optimizer are MAE and Adam optimizer~\cite{adam_ref}, respectively. Hyperparameter tuning was performed on the learning rate, batch size, number of cells, and layers to arrive at the lowest MAE score. 
\section{Evaluation}
\label{sec:evaluation}
This section introduces the test methodology used in this paper and then discusses the experimental results.
\subsection{Evaluation Setup}
We use \SI{75}{\percent} of the thousand UHD video sequences of the Inter4K Dataset~\cite{inter4k_ref} as the training dataset and set \SI{5}{\percent} as the validation set. The remaining \SI{20}{\percent} are set as the test dataset. All experiments are run on a system with an Intel i7-11370H processor and 16GB RAM. All video sequences are encoded using the x264 AVC encoder with CRF values between \SI{1}{} and \SI{51}{} to induce different quality distortions, and the corresponding VMAF is evaluated. Each segment comprises eight frames, \ie $f_{c}=8$. Hence, a video sequence is divided into 15 segments. The original and reconstructed video segments' luma texture features are extracted with the VCA v2.0 open-source video complexity analyzer running with eight CPU threads. Furthermore, the original and reconstructed video feature extraction process is implemented concurrently, with four CPU threads for each process. The ground truth VMAF scores are computed using the model released by Netflix~\cite{vmaf_model_ref}. Pearson correlation coefficient (PCC) and Mean Absolute Error (MAE) scores are analyzed between the \rrtif scores and the ground truth VMAF quality scores.   In addition, $\tau_{\text{T}}$ and $J_{\text{T}}$, \ie the total time taken and energy consumed to compute the quality metrics, are evaluated. We used \textit{codecarbon} software to measure the energy consumption~\cite{codecarbon_ref}.
\subsection{Relevance of features}
The importance of the input of features to the LSTM-based fusion model is analyzed using the univariate approach. All the other feature values are set to zero, the MAE is computed, and this is subtracted from the MAE of the model with all features intact, which gives a measure of the decrease in accuracy (\ie increase in error) when that feature is removed from the model~\cite{feature_imp}. Subsequently, the absolute value of the decrease in accuracy is computed and normalized to obtain the importance score, where higher absolute scores indicate more critical features. The information factor is calculated by applying Min-Max normalization to the average MAE scores. The importance of the features is visualized in Figure~\ref{fig:rrtif_feat_imp}. It is observed that the SSIM feature contributes the most to the \rrtif estimation, followed by $r_{\text{E}}$, $r_{\text{h}}$, and $r_{\text{L}}$ features.
\subsection{Accuracy}
\begin{figure}[t]
\centering
\begin{subfigure}{0.47\columnwidth}
    \centering
    \includegraphics[width=\textwidth]{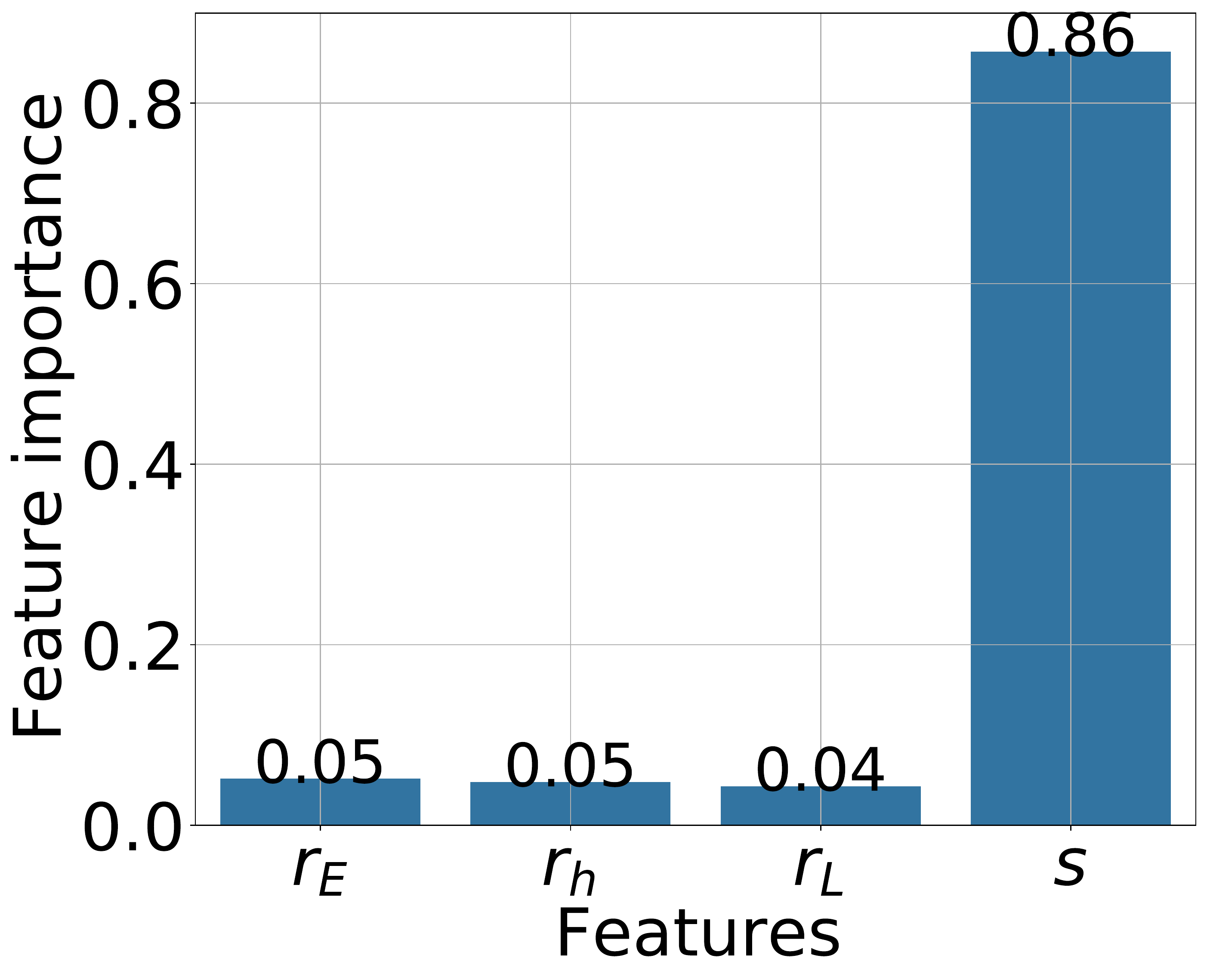}
    \caption{Univariate feature importance of the LSTM model}
        \label{fig:rrtif_feat_imp}
\end{subfigure}
\hfill
\begin{subfigure}{0.47\columnwidth}
    \centering
    \includegraphics[width=\textwidth]{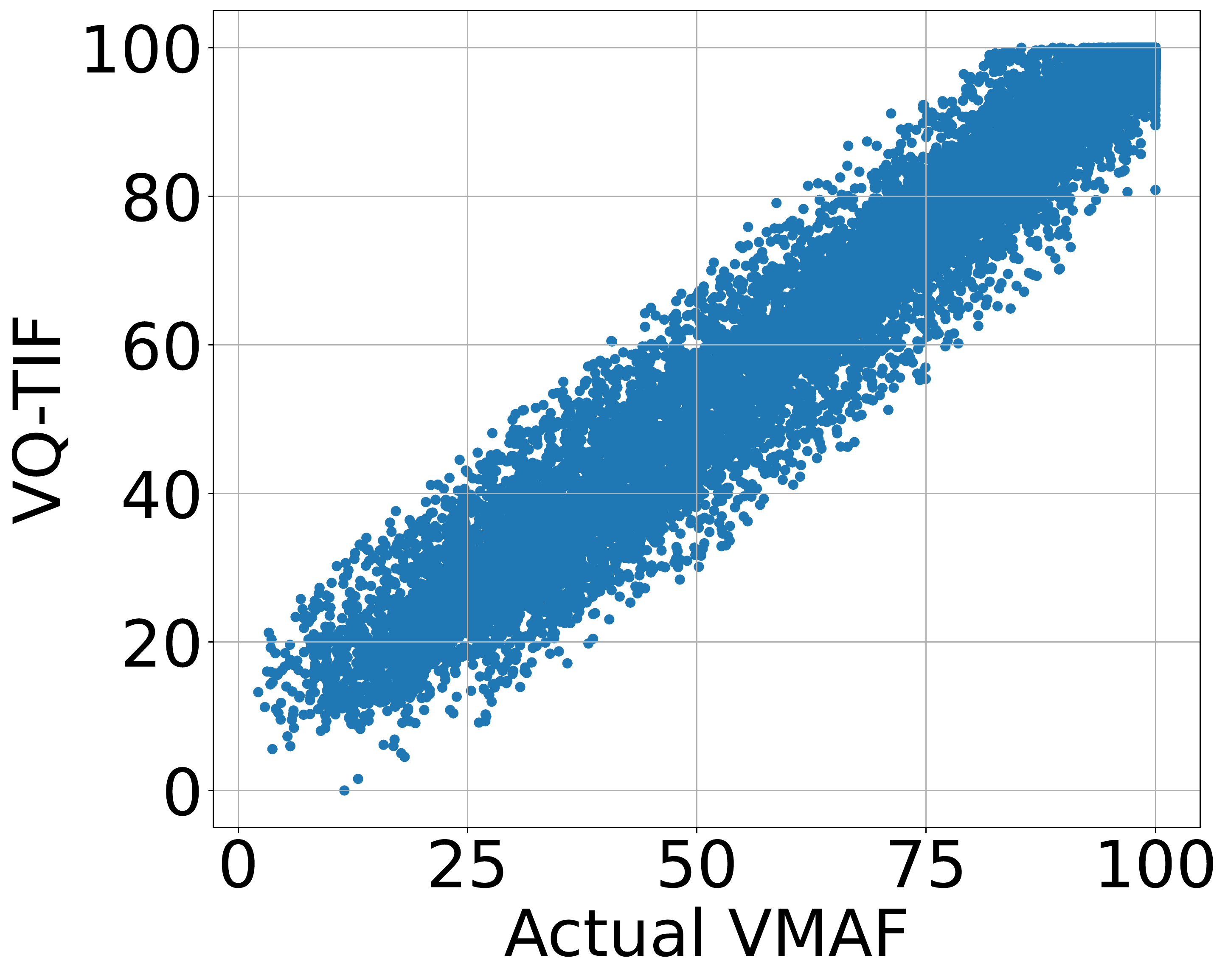}
    \caption{Scatterplot of VMAF and \rrtif}
        \label{fig:rrtif_scatter_plot}
\end{subfigure}
\hfill
\begin{subfigure}{0.47\columnwidth}
    \centering
    \includegraphics[width=\textwidth]{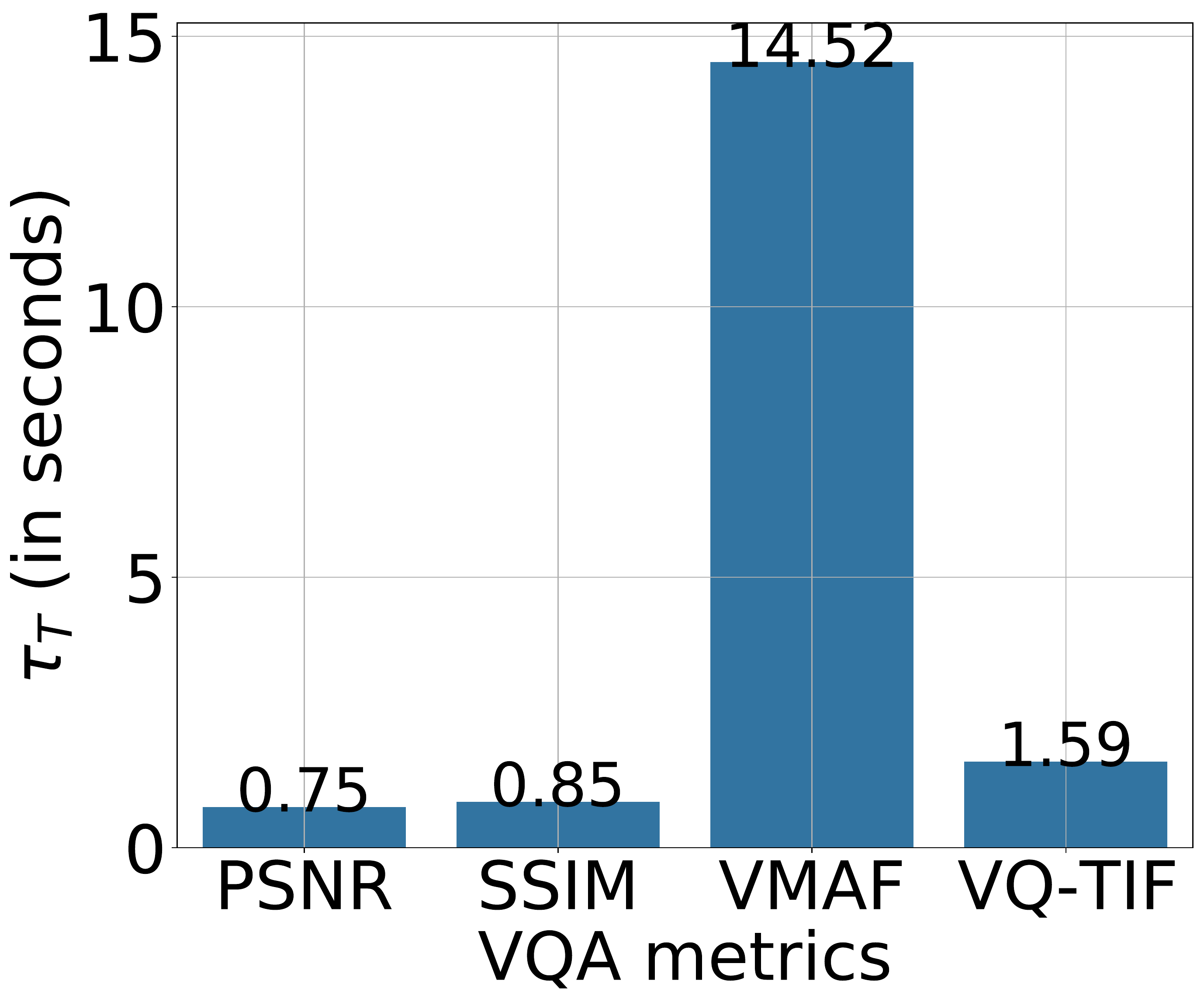}
    \caption{Total processing time ($\tau_{T}$)}
        \label{fig:rrtif_time_plot}
\end{subfigure}
\hfill
\begin{subfigure}{0.47\columnwidth}
    \centering
    \includegraphics[width=\textwidth]{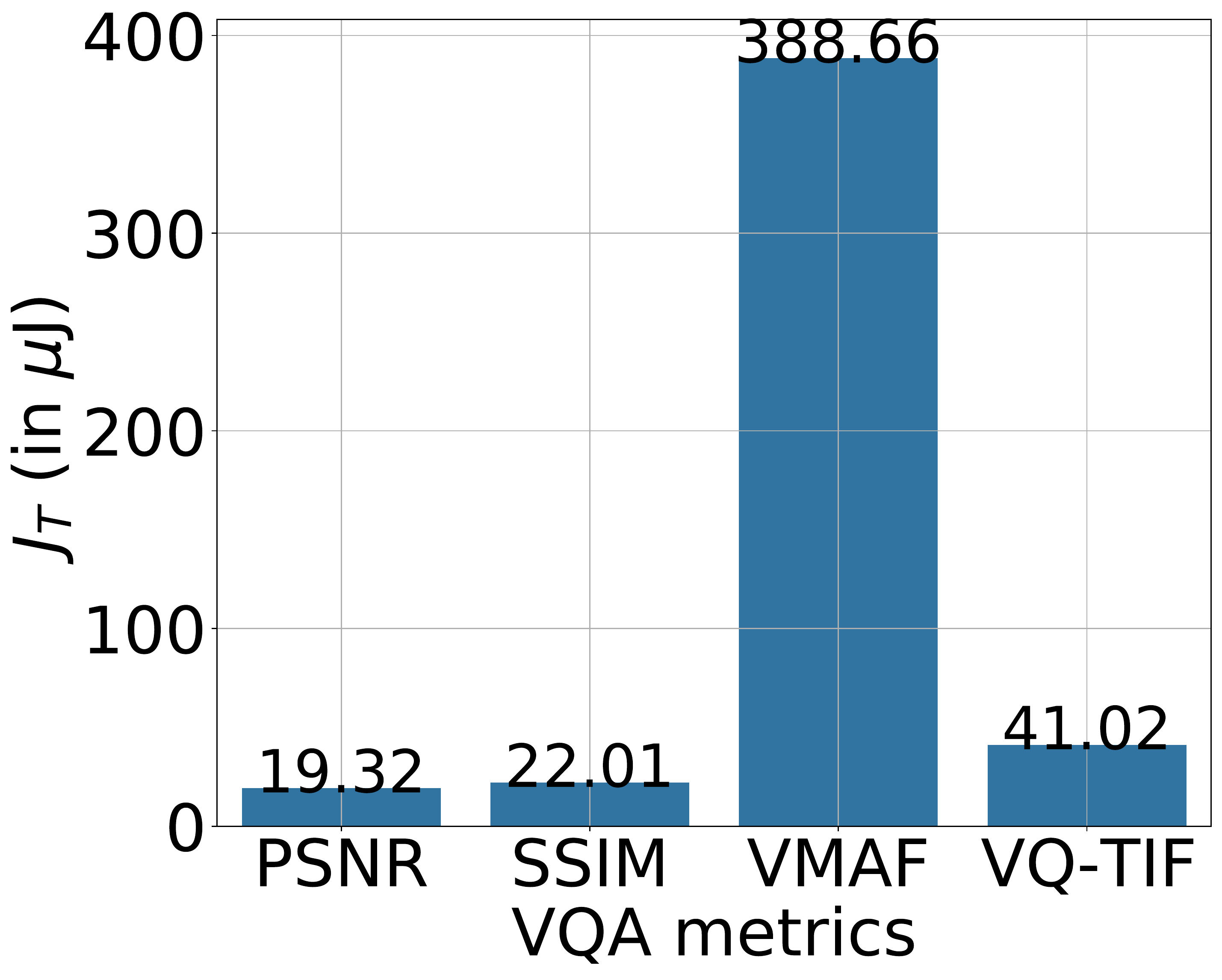}
    \caption{Total energy consumption ($J_{T}$)}
    \label{fig:rrtif_energy_plot}    
\end{subfigure}
\caption{Prediction results.}
\label{fig:rrtif_time_energy_plot}
\end{figure}

We evaluated the precision of \rrtif using \pcc~(PCC)~\cite{pcc_ref} between the VMAF and \rrtif scores.
Figure~\ref{fig:rrtif_scatter_plot} shows the scatterplot of the \rrtif scores and the VMAF scores. A strong correlation between scores is observed. The average PCC of the \rrtif scores to the VMAF score in the evaluation dataset is \SI{0.96}{}, while MAE is \SI{2.71}{}. The maximum deviation between the \rrtif and VMAF score is 20.23 points. The accuracy can be improved further by reducing $f_c$, increasing the inference time.  

\subsection{Processing time and energy}
We observe the computation time and energy consumed for texture information extraction in \rrtif as \SI{0.67}{\second} (\ie 179.10\,fps) and \qty{17.18}{\uJ}, respectively (\cf Figure~\ref{fig:rrtif_time_plot} and \ref{fig:rrtif_energy_plot}). The time taken and energy consumed for the SSIM computation are \SI{0.85}{\second} and \qty{22.01}{\uJ}, respectively. The time taken and the energy consumed for texture information fusion are \SI{0.07}{\second} and \qty{1.83}{\uJ}, respectively. Therefore, the total processing time, $\tau_{\text{T}}$ is \SI{1.59}{\second} (\ie 75.47\,fps), while the processing time for the state-of-the-art VMAF computation is \SI{14.52}{\second} (\ie 8.26\,fps). The computation speed of \rrtif is 9.14 times higher than the state-of-the-art VMAF evaluation. Furthermore, in terms of total energy consumption, \rrtif saves \SI{89.44}{\percent} compared to the state-of-the-art VMAF implementation.
\section{Conclusions}
\label{sec:conclusion_future_dir}
We proposed \rrtif, a fast and accurate reduced-reference video quality assessment (RR-VQA) method based on texture information fusion. \rrtif includes DCT-energy-based video complexity feature extraction where features representing luma texture and temporal activity are extracted from the original and reconstructed video segments. The extracted texture information is fused using an LSTM-based model to determine the \rrtif score. It is observed that \rrtif is determined at a speed of \SI{9.14}{} times faster than the state-of-the-art implementation of VMAF for Ultra HD (2160p) videos, consuming \SI{89.44}{\percent} less energy. At the same time, \rrtif scores yield a PCC of \SI{0.96}{} and MAE of \SI{2.71}{} compared to the VMAF scores.

The evaluation of the proposed \rrtif model is limited to static dynamic range (SDR) content. The evaluations on high dynamic range (HDR) content and the associated optimizations are subject to future work. Furthermore, the \rrtif model can be extended to determine visual quality at multiple resolutions, including 8K (4320p). Furthermore, various signal distortions may be considered during the model's training to enhance the application scope.

\section{Acknowledgment}
The financial support of the Austrian Federal Ministry for Digital and Economic Affairs, the National Foundation for Research, Technology and Development, and the Christian Doppler Research Association is gratefully acknowledged. Christian Doppler Laboratory ATHENA: \url{https://athena.itec.aau.at/}.
\balance
\bibliographystyle{IEEEtran}
\bibliography{references.bib}
\balance
\end{document}